\documentclass[8.5pt,twoside,twocolumn]{article}
\usepackage[super,sort&compress,comma]{natbib} 
\usepackage{times,mathptmx}
\usepackage{sectsty}
\usepackage{balance} 

\usepackage{graphicx} 
\usepackage{lastpage}
\usepackage[format=plain,justification=raggedright,singlelinecheck=false,font=small,labelfont=bf,labelsep=space]{caption} 
\usepackage{fancyhdr}
\begin{document}

\thispagestyle{plain}
\fancypagestyle{plain}{
\fancyhead[L]{\includegraphics[height=8pt]{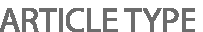}}
\fancyhead[C]{\hspace{-1cm}\includegraphics[height=20pt]{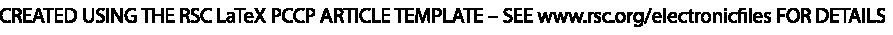}}
\fancyhead[R]{\includegraphics[height=10pt]{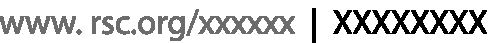}\vspace{-0.2cm}}
\renewcommand{\headrulewidth}{1pt}}
\renewcommand{\thefootnote}{\fnsymbol{footnote}}
\renewcommand\footnoterule{\vspace*{1pt}%
\hrule width 3.4in height 0.4pt \vspace*{5pt}} 
\setcounter{secnumdepth}{5}

\makeatletter 
\def\subsubsection{\@startsection{subsubsection}{3}{10pt}{-1.25ex plus -1ex minus -.1ex}{0ex plus 0ex}{\normalsize\bf}} 
\def\paragraph{\@startsection{paragraph}{4}{10pt}{-1.25ex plus -1ex minus -.1ex}{0ex plus 0ex}{\normalsize\textit}} 
\renewcommand\@biblabel[1]{#1}            
\renewcommand\@makefntext[1]%
{\noindent\makebox[0pt][r]{\@thefnmark\,}#1}
\makeatother 
\renewcommand{\figurename}{\small{Fig.}~}
\sectionfont{\large}
\subsectionfont{\normalsize} 

\fancyfoot{}
\fancyfoot[LO,RE]{\vspace{-7pt}\includegraphics[height=9pt]{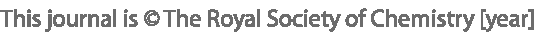}}
\fancyfoot[CO]{\vspace{-7.2pt}\hspace{12.2cm}\includegraphics{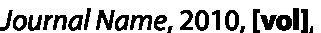}}
\fancyfoot[CE]{\vspace{-7.5pt}\hspace{-13.5cm}\includegraphics{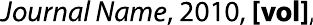}}
\fancyfoot[RO]{\footnotesize{\sffamily{1--\pageref{LastPage} ~\textbar  \hspace{2pt}\thepage}}}
\fancyfoot[LE]{\footnotesize{\sffamily{\thepage~\textbar\hspace{3.45cm} 1--\pageref{LastPage}}}}
\fancyhead{}
\renewcommand{\headrulewidth}{1pt} 
\renewcommand{\footrulewidth}{1pt}
\setlength{\arrayrulewidth}{1pt}
\setlength{\columnsep}{6.5mm}
\setlength\bibsep{1pt}

\twocolumn[
  \begin{@twocolumnfalse}
\noindent\LARGE{\textbf{Enhanced diffusion of tracer particles in
    dilute bacterial suspensions}}
\vspace{0.6cm}

\noindent\large{\textbf{Alexander Morozov$^{\ast}$ and Davide Marenduzzo }}\vspace{0.5cm}

\noindent\textit{\small{\textbf{Received Xth XXXXXXXXXX 20XX, Accepted Xth XXXXXXXXX 20XX\newline
First published on the web Xth XXXXXXXXXX 200X}}}

\noindent \textbf{\small{DOI: 10.1039/b000000x}}
\vspace{0.6cm}

\noindent \normalsize{

Swimming bacteria create long-range velocity fields that stir a large volume of fluid and move around passive particles dispersed in the fluid. Recent experiments and simulations have shown that long-time mean-squared displacement of passive particles in a bath of swimming bacteria exhibits diffusive behaviour with the effective diffusion coefficient significantly larger than its thermal counterpart. Comprehensive theoretical prediction of this effective diffusion coefficient and understanding of the enhancement mechanism remain a challenge. Here, we adapt the kinetic theory by Lin \textit{et al., J. Fluid Mech.}, 2011, \textbf{669}, 167 developed for 'squirmers' to the bacterial case to \emph{quantitatively} predict enhanced diffusivity of tracer particles in dilute two- and three-dimensional suspensions of swimming bacteria. We demonstrate that the effective diffusion coefficient is a product of the bacterial number density, their swimming speed, a geometric factor characterising the velocity field created by a single bacterium, and a numerical factor. We show that the numerical factor is, in fact, a rather strong function of the system parameters, most notably the run length of the bacteria, and that these dependencies have to be taken into account to quantitatively predict the enhanced diffusivity. We perform molecular-dynamics-type simulations to confirm the conclusions of the kinetic theory. Our results are in a good agreement with the values of enhanced diffusivity measured in recent two- and three-dimensional experiments.}
\vspace{0.5cm}
 \end{@twocolumnfalse}
  ]

\section{Introduction}
\footnotetext{\textit{~SUPA, School of Physics and Astronomy,
    University of Edinburgh, Edinburgh, UK. E-mail: Alexander.Morozov@ph.ed.ac.uk}}

Recent interest in suspensions of self-propelled colloidal particles steams from their relevance to a variety of disciplines\cite{Ramaswamy2010}. In physics, they provide one of the simplest model to understand statistical mechanics of out-of-equilibrium systems \cite{Cates2012review} and hydrodynamics and rheology of active matter\cite{LaugaPowers,Marchetti2013}. In biology, the motility of bacteria and eukaryotic microorganisms is linked to understanding various diseases\cite{Celli2009}, fertility\cite{sperm} and biomixing in oceans\cite{Katija2012}. In engineering, it has been demonstrated that motile particles can be made to perform work\cite{Angelani2009,DiLeonardo2010,Sokolov2010} and deliver cargo\cite{cargo}. 

Bacteria are one of the most readily available realisations of self-propelled particles. Their individual motility and collective behaviour have been extensively studied\cite{bergbook,Dombrowski2004}. Many species propel by pushing the surrounding fluid backwards by rotating long thin flagella. The propulsive force applied to the fluid is then compensated by the drag the fluid exerts on the bacterium. Thus, locally, bacteria act as self-propelled force-dipoles that stir the fluid in a large volume around them. The long-ranged velocity fields created by bacteria result in an induced motion of passive particles suspended in the fluid such as dead bacteria, nutrients, small droplets of other fluids \emph{etc.} This so-called \emph{enhanced diffusion} is potentially relevant for inducing feeding currents around microorganisms and biomixing in oceans\cite{Katija2012}.

Systematic study of enhanced diffusion started with the pioneering work by Wu and Libchaber\cite{wu2000}, who measured the effective diffusion coefficient of large colloidal particles in an \emph{E. coli} suspension in a quasi-2D free standing soap film. Wu and Libchaber\cite{wu2000} concluded that at long times colloidal particles behaved diffusively with the effective diffusion coefficient being about $100$ times larger than the thermal one. Since then many studies have confirmed similar behaviour. Long-time diffusive behaviour of tracers was observed in dilute suspensions of \emph{E. coli}\cite{Kim2004,chen2007,mino2011,Wilson2011,mino2013,alys2013}, \emph{B. subtilis}\cite{Chantal2011}, alga \emph{Chlamydomonas reinhardtii}\cite{Leptos2009,kurtuldu2011}, and synthetic swimmers\cite{mino2011}, with experiments performed in quasi-2D thin films\cite{wu2000,mino2011,mino2013,Chantal2011,kurtuldu2011} or 3D geometries\cite{Kim2004,chen2007,Leptos2009,Wilson2011,alys2013}. These studies employed either colloidal particles or non-motile bacteria as tracers, both comparable in size with the swimmers, with the exception of the work by Kim and Breuer\cite{Kim2004}, who considered diffusion of small \emph{Dextran} molecules in a bath of \emph{E. coli} bacteria. On the theoretical side, simulations of tracers with self-propelled particles of various types confirm diffusive behaviour\cite{Underhill2008,Ishikawa2010,childress2011} with a diffusion coefficient significantly larger than its equilibrium value in the absence of swimmers. 

Experiments\cite{Leptos2009,mino2011,mino2013,alys2013}, theory\cite{Underhill2008,Childress2010,childress2011,pushkin2013jfm,mino2013,alys2013,pushkin2013prl}, and simulations\cite{Ishikawa2010,childress2011} provide evidence that the enhanced diffusion coefficient scales linearly with the so-called \emph{active flux}: the product of the number density of swimmers $n$ and their swimming speed $U$. In order to obtain a quantity of the same dimensions as the diffusion coefficient, the active flux should be multiplied by a lengthscale to the fourth power. The precise understanding of the origin of this lengthscale and the value of the numerical prefactor in the scaling relation is a subject of active ongoing research. Childress and co-workers\cite{Childress2010,childress2011} have developed a kinetic theory for enhanced diffusion by spherical "squirmers" performing run-and-tumbling motion. Their theory is based on far-field hydrodynamic interactions between swimmers and tracers. It confirms the linear scaling with the active flux, and identifies the size of the squirmer as the scaling lengthscale. Their result is in a good quantitative agreement with the measurements by Leptos \emph{et al.}\cite{Leptos2009} provided the strength of the squirmer velocity field was selected appropriately. A similar theory was later developed by Mi\~no \emph{et al.}\cite{mino2013} for \emph{E. coli} swimming in the bulk and next to solid surfaces. They assumed that the run length of the swimming bacteria is, essentially, infinite and obtained a prediction for the effective diffusion coefficient of a tracer that was significantly below their measured values. That theory was later modified to include realistic run length of the \emph{E. coli} bacteria by Jepson \emph{et al.}\cite{alys2013} and the resulting prediction is in a very good agreement with their 3D measurements. Recently, Pushkin and co-workers advocated a different mechanism for enhanced diffusivity of tracers based on the ideas of entrainment\cite{pushkin2013jfm,pushkin2013prl}. They argue that large tracer displacements can be caused by close encounters between the tracer and a bacterium, with the former travelling in the hydrodynamic wake of the latter. For bacteria, this mechanism is claimed to be especially important in 2D.

The present paper essentially advocates the mechanism introduced by Childress and co-workers \cite{Childress2010,childress2011} and is a development of our simple theory used to explain the 3D measurements by Jepson \emph{et al.}\cite{alys2013} We show that while the scaling form of the diffusion coefficient discussed above is a good first approximation, the numerical prefactor, which was assumed to be constant by previous studies\cite{Childress2010,childress2011,mino2013,alys2013,pushkin2013prl}, in fact depends on the properties of the swimmers, and these dependencies should be considered if one is to predict experimental results quantitatively. We also discuss the effect of the near-field of the swimmers and demonstrate that it is largely irrelevant in our model. 

Our paper is organised as follows. In Section \ref{theory} we introduce our semi-analytical model and estimate the 3D effective diffusion coefficient of a tracer in a dilute \emph{E. coli} suspension. In Section \ref{simulations} we perform direct numerical simulations of a tracer particle immersed into a bath of dipole-like swimmers and compare our results with the estimate of Section \ref{theory}. We study in detail how the effective diffusion coefficient depends on the properties of the swimmers beyond the scaling relation discussed above. In Section \ref{2Dtheory} we repeat the calculation of Section \ref{theory} for a 2D suspension next to a solid boundary and discuss how this case differs from the 3D enhanced diffusion. We also calculate how the effective diffusion coefficient changes with distance to a wall. Finally, we discuss the limitations of our results and their implications for the study of transport in bacterial suspension.

\section{Theory}
\label{theory}

Here we present a semi-analytical method to estimate the enhanced diffusivity of a tracer in a 3D bath of swimming bacteria. We consider a dilute solution of bacteria at number density $n$, which is typically $n \sim 10^{-3} \mu m^{-3}$ in the recent experiments\cite{Kim2004,mino2013,alys2013}. We assume that the bacteria perform a simplified version of run-and-tumble motion: they swim in a straight line with a constant speed $U$ and then instantaneously randomly change their swimming direction. The distance $\lambda$ travelled between two reorientations is fixed. Wild-type \emph{E. coli} bacteria have a distribution of the run length $\lambda$ and can vary the properties of this distribution (mean, width, \emph{etc.}) to adapt to local gradients of oxygen, nutrients, light \emph{etc.}\cite{bergbook}, but we ignore this for simplicity. The size of the tracer $R_0$ is considered to be sufficiently small compared to the typical distance between two bacteria. 

In the absence of hydrodynamic interactions, the tracer particle would only move due to direct collisions with the swimming bacteria\cite{Chantal2011}. The corresponding effective diffusion coefficient of the tracer can be estimated based on a simple argument. In 3D, the probability that a bacterium can hit the tracer is given by the portion of its "horizon" blocked by the tracer, $\pi R_0^2/{4\pi r^2}$, where $r<\lambda$ is the distance between the bacterium and the tracer. After a time $t$, the number of bacteria hitting the tracer from the shell $(r,r+dr)$ is given by the product of this probability, the volume of the shell $4\pi r^2 dr$, the number density of the bacteria $n$, and the number of "runs" performed by one bacterium in that time interval $U t/\lambda$, since the latter gives the number of scattering attempts per bacterium in time $t$. Since the bacteria can only hit the tracer swimming in a straight line from inside the sphere of radius $\lambda$, the total number of collisions in time $t$ is given by the integral of this expression over $r$ from $0$ to $\lambda$. Finally, we assume that in each collision the tracer is displaced by a typical distance $R_b$ which is comparable to the size of the bacterium, in other words, it is pushed aside by a swimming bacterium. The effective diffusion coefficient $D_c$ due to direct collisions is then defined by $6 D_{c} t \equiv n U t \pi R_0^2 R_b^2$. From the experiments by Jepson \emph{et al.} \cite{alys2013}, $U\sim 15 \mu m/s$, $R_0\sim R_b\sim1\mu m$ and $n\sim 10^{-3}\mu m^{-3}$, giving $D_c\sim 10^{-2} \mu m^2/s$, while the experimentally observed enhancement of the diffusion coefficient is $\Delta D\sim 10^{-1} \mu m^2/s$. Although only a relatively small adjustment of $R_0$ and $R_b$, for example, is required to obtain an estimate of $D_c$ similar to the measured value, we note here that that would be a rather meaningless coincidence: in a fluid a direct collision is impossible because the tracer would start moving away from the approaching bacterium\cite{Dunkel2010} long before their separation is of order $R_0+R_b$. Therefore, a proper estimate of the effective diffusion coefficient should take into account the long-range nature of the hydrodynamic interactions and consider displacements of the tracer due to bacteria moving anywhere in the system, not only the bacteria that collide directly with the tracer as in the estimate above. However, the derivation of $D_c$ demonstrates the origin of the linear scaling of the enhanced diffusion coefficient with the active flux $n U$. In what follows we show that hydrodynamic interactions preserve the linear scaling $D\sim n U$ but, additionally, introduce subtle dependencies on the properties of the swimmers. 

To model \emph{E. coli} bacteria, we assume that each swimmer creates a dipolar velocity field \cite{kimkarrila,Spagnolie2012}
\begin{equation}
{\bf u}\left( {\bf r}\right) = \frac{p\,{\bf r}}{r^3} \left( 3\cos^2{\theta} - 1\right),
\label{dipolar}
\end{equation}
where $\bf r$ is the radius-vector of the observation point w.r.t. the swimmer, $|{\bf r}|=r$, $\theta$ is the angle between the direction of swimming and $\bf r$, and $p$ sets the strength of the dipolar field. Drescher \emph{et al.} \cite{Drescher2011} have measured the velocity field created by a single \emph{E. coli} bacterium far away from surfaces and confirmed that its far-field contribution is reasonably described by Eq.(\ref{dipolar}) with $p=31.8 \mu m^3/s$ for bacteria swimming with $U=22\mu m/s$. While \emph{E. coli} bacteria produce additional velocity field close to its body as measured by Drescher \emph{et al.} \cite{Drescher2011}, the influence of this near-field on the motion of the tracer will only be significant at very small bacterium-tracer separations. During such close encounters, motion of the tracer and the bacterium is dominated by lubrication, electrostatic, van der Waals, \emph{etc.} forces and is rather complicated. However, since in dilute suspensions close encounters should be relatively rare, we will be ignoring the effect of the near-field and other forces potentially relevant at small separations. The relative importance of the near-field effects will be further assessed in Section \ref{discussion}.

The low density regime chosen here allows us to make further important assumptions. First, the total velocity field created by the swimming bacteria is assumed to be a linear superposition of the velocity fields of individual swimmers, i.e. we ignore hydrodynamic interactions between bacteria. Second, we assume that the effective diffusion coefficient of the tracer particle does not depend on the tracer size. According to Fax\'{e}n law \cite{kimkarrila}, the force on a spherical tracer of size $R_0$ immersed into an externally-generated velocity field ${\bf v_{\infty}}({\bf r})$ is given by 
\begin{equation}
{\bf F} = 6\pi\eta R_0 \left( 1+\frac{R_0^2}{6}\nabla^2\right){\bf v_{\infty}}({\bf r}) |_{\rm surf} - 6\pi\eta R_0 {\bf U},
\label{faxen}
\end{equation}
where ${\bf U}$ is the velocity of the tracer, and the first term is evaluated at the surface of the sphere. In the absence of external forces acting on the tracer, ${\bf F} = 0$. When the velocity field does not significantly change over the distance comparable with $R_0$, the derivative term in Eq.(\ref{faxen}) can be neglected, and Fax\'{e}n law predicts that the sphere will move with the velocity of the externally generated velocity field at its location, ${\bf U}={\bf v_{\infty}}({\bf r}) |_{\rm surf}$, i.e. the sphere will behave as a passive tracer. This regime breaks down when the external velocity field rapidly changes on the scale of $R_0$. In our problem the external field ${\bf v_{\infty}}$ is generated by swimming point-like dipoles with the velocity fields given by Eq.(\ref{dipolar}). Using $v_{\infty}\sim r^{-2}$, we conclude that the sphere is a passive tracer as long as $R_0^2/r^2\ll1$, where $r$ is the typical tracer-swimmer separation. Therefore our assumption that the tracer particle is being passively advected by the velocity field created by the swimmers is justified for low number density of the swimmers. In the recent experiments of Jepson \emph{et al.}\cite{alys2013},  $R_0^2/r^2\sim 10^{-3}$ when they used dead bacteria without flagella as tracers. Jepson \emph{et al.}\cite{alys2013}, and also Mi\~{n}o and co-workers\cite{mino2011,mino2013}, have confirmed that the enhanced diffusion coefficient was independent of the tracer size. Therefore, in what follows we assume that the tracer is a point-like particle.

\begin{figure}[t]
\includegraphics[width=12cm]{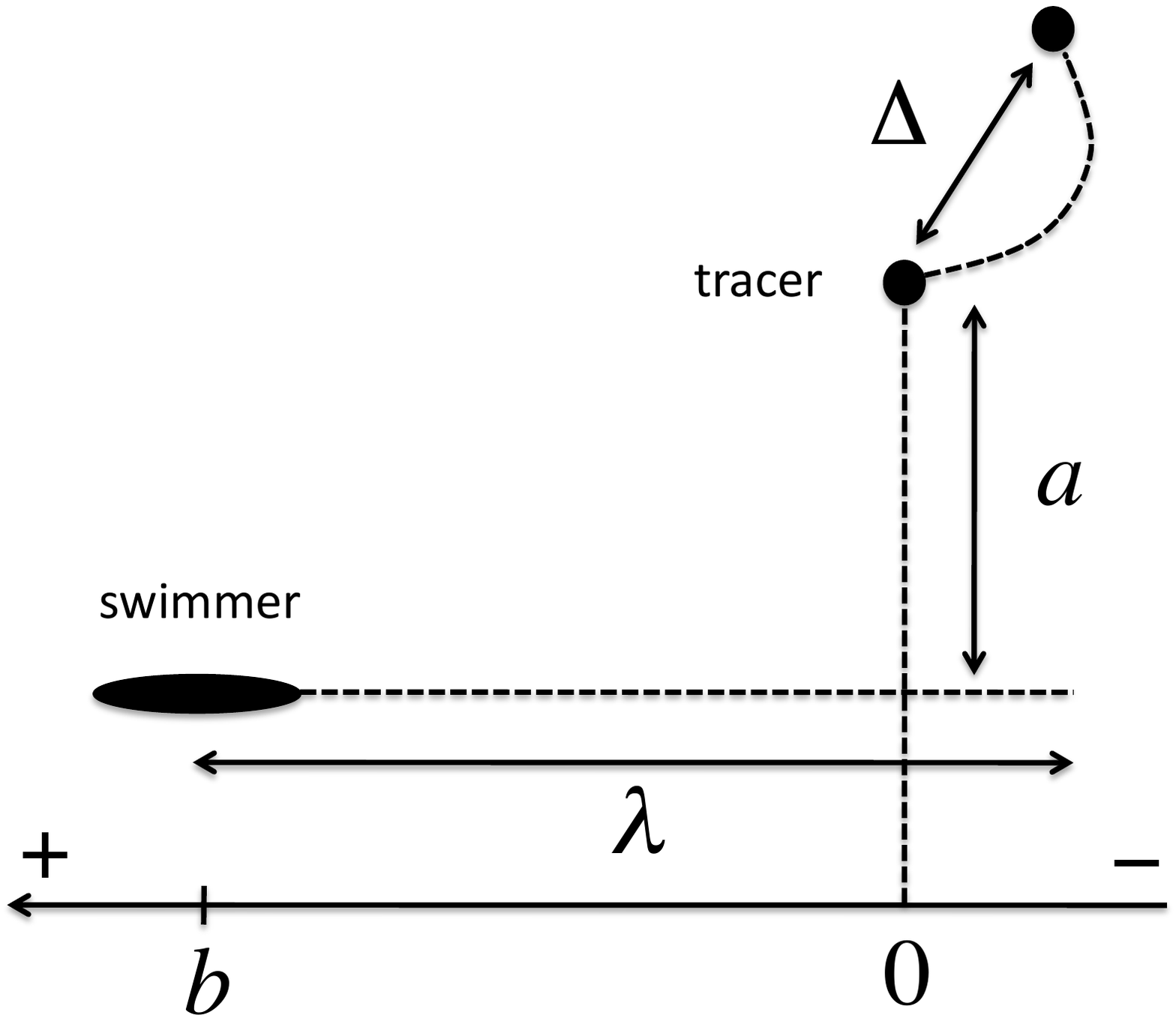}
\caption{Schematics of a scattering event between a swimmer and a tracer: $a$ is the shortest distance between the path of the swimmer and the original position of the tracer, and $b$ is the distance between the original position of the swimmer and the point of the closest approach. The total distance travelled by the swimmer is $\lambda$ and $\Delta$ denotes the net displacement of the tracer during the scattering event.}
\label{event}
\end{figure}

Our calculation is based on the kinetic theory developed for "squirmers" by Lin \emph{et al.} \cite{childress2011}. Here we adopt their method for the case of bacteria. The main ingredient of the theory is the net displacement $\Delta$ of the tracer particle by a swimming bacterium from the moment the latter had acquired a particular swimming direction and started moving in a straight line until it has travelled the distance $\lambda$. Each such scattering event can be parameterised by two lengths: the shortest distance $a$ between the path of the swimmer and the original position of the tracer, and the distance $b$ between the original position of the swimmer and the point of the closest approach; see Fig.\ref{event}. The theory by Lin \emph{et al.} \cite{childress2011} is based on the assumption that the net mean-squared displacement $\langle |\Delta{\bf r}(t)|^2\rangle$ of the tracer particle after time $t$ can be viewed as a superposition of individual scattering events, and, therefore, it can be approximated by
\begin{equation}
\langle |\Delta{\bf r}(t)|^2\rangle = M(t) \langle \Delta(a,b)^2 \rangle_{a,b},
\label{eq3}
\end{equation}
where $M(t)$ is the number of scattering events during time $t$, $\Delta(a,b)$ is the net displacement of the tracer during a scattering event with the initial parameters $a$ and $b$, and $\langle\dots\rangle_{a,b}$ denotes averaging over all possible scattering configurations. Lin \emph{et al.} \cite{childress2011} have shown that Eq.(\ref{eq3}) can be rewritten as
\begin{equation}
\langle |\Delta{\bf r}(t)|^2\rangle = n \frac{U t}{\lambda}  \int_{0}^{\infty}da \int_{-\infty}^{\infty} db\,2\pi\,a\,\Delta^2(a,b).
\label{int_original}
\end{equation}
The prefactor in Eq.(\ref{int_original}) gives the effective density of the swimmers: during time $t$, each swimmer changes its swimming directions $U\, t/\lambda$ times which implies that during that time period there will be $n_A\,U\,t/\lambda$ scattering events of the type shown in Fig.\ref{event} per unit volume. The average of $\Delta^2$ over all possible positions and orientations of the swimmer is written as an integral over the whole space in cylindrical coordinates with $a$ and $b$ along the radial and axial directions, respectively. The effective diffusion coefficient of the tracer $D$ is then defined through $\langle |\Delta{\bf r}(t)|^2\rangle=6\,D\,t$. 

To aid evaluation of Eq.(\ref{int_original}), we introduce a new quantity $\sigma = \sqrt{p/U}$ based on the dipolar strength $p$ of the swimmer, Eq.(\ref{dipolar}), and its propulsion speed $U$. This quantity has dimensions of length but should not be understood as a single lengthscale; we will discuss this in more detail in Section \ref{discussion}. We introduce new variables, $a = \sigma e^{\xi}$, $b=\lambda \chi$ and $\Delta = \sigma \tilde\Delta$, and obtain the final expression for the effective diffusion coefficient
\begin{equation}
D = A\,n\,U \sigma^4 = A\,n\,U \left( \frac{p}{U}\right)^2,
\label{main}
\end{equation}
where
\begin{equation}
A = \frac{\pi}{3}\int_{-\infty}^{\infty}d\xi \int_{-\infty}^{\infty}d\chi\,e^{2\xi}\,\tilde\Delta^2(\xi,\chi).
\label{A}
\end{equation}
Eq.(\ref{main}) is similar to the results obtained by previous authors\cite{Underhill2008,childress2011,mino2013,pushkin2013prl} in that it predicts that the effective diffusion coefficient scales linearly with the active flux $n\,U$ and the fourth power of a lengthscale. While $A$ appears to be a numerical prefactor, we will demonstrate below that it is, actually, a weak function of $p$, $U$, $n$ and other parameters, and a strong function of $\lambda$. 

To evaluate $\tilde\Delta(\xi,\chi)$ from Eq.(\ref{A}), we integrate numerically the equations of motion for the tracer, $\dot{\bf r}_t(t) = {\bf u}\left( {\bf r}_t(t) - {\bf r}_s(t)\right)$ using an explicit Euler time-iteration scheme \cite{hussainibook} with a timestep $dt$. Here, ${\bf r}_t(t)$ and ${\bf r}_s(t) = {\bf r}_s(0) + U\,t\,{\bf e}$ are the positions of the tracer and swimmer, respectively, and the velocity field ${\bf u}$ is given by Eq.(\ref{dipolar}). At $t=0$, the tracer is at the origin. The initial position of the swimmer and the direction of swimming ${\bf e}$ are set by the scattering parameters $a$ and $b$. To avoid numerical problems related to the singular nature of the velocity field in Eq.(\ref{dipolar}), we introduce a cut-off length $c$: at each timestep the tracer is moved only if $\left|{\bf r}_t(t) - {\bf r}_s(t)\right|>c$, otherwise only the position of the swimmer is updated. The precise physical meaning of $c$ and the influence of its choice on $A$ will be discussed below.

\begin{figure}[t]
\begin{center}
\includegraphics[width=9.cm]{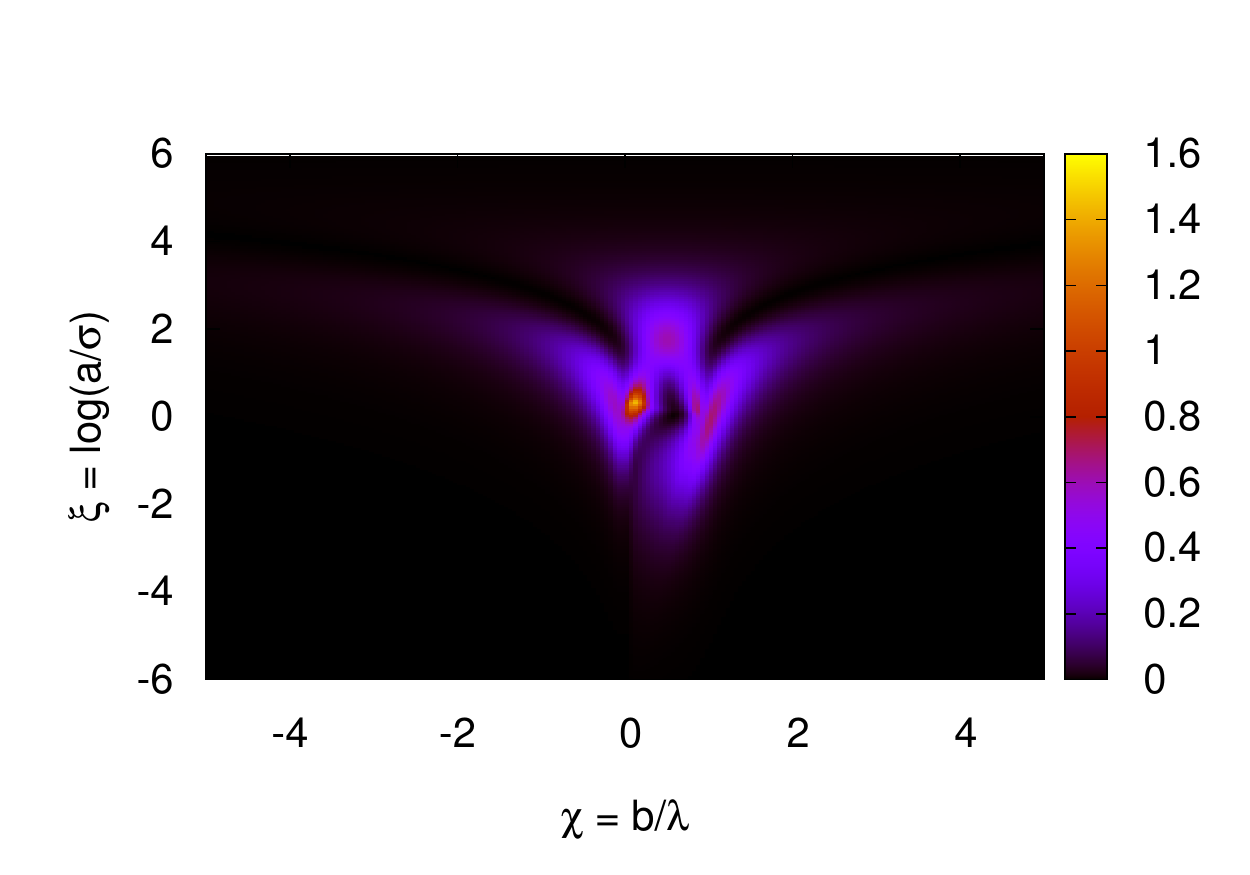}
\end{center}
\caption{The value of the integrand $e^{2\xi}\,\tilde\Delta^2(\xi,\chi)$ from Eq.(\ref{A}) for $p=32$, $U=22$, $c=1$, $dt=0.001$, $\lambda=10$.}
\label{colourmap}
\end{figure}

Our choice of the numerical values of parameters is motivated by the experiments of Drescher \emph{et al.} \cite{Drescher2011} who measured the velocity field around a freely swimming \emph{E. coli} bacterium and obtained $p=31.8\mu m^3/s$ for the bacterium swimming with the average speed $U=22\mu m/s$. In the rest of the paper we adopt a system of units where the length is measured in micrometers and time in seconds. In these units, we set $p=32$, $U=22$, and the run length $\lambda=10$, unless stated otherwise. Since we are mainly interested in the enhanced diffusion of large colloidal particles, the minimal bacterium-tracer separation is comparable to the size of the bacterium and we set $c=1$. We checked that for any combination of parameters time-iteration converges for $dt<0.005$, and choose $dt=0.001$.

\begin{figure}[t]
\begin{center}
\includegraphics[width=8.cm]{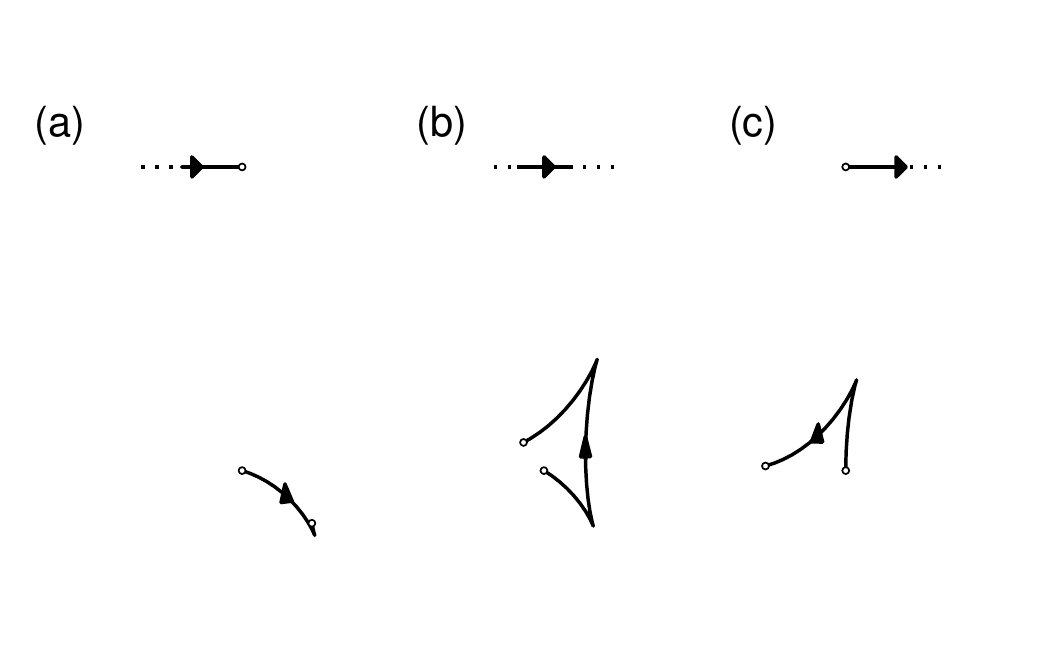}
\end{center}
\caption{Scattering of a tracer by a swimmer with $a=1.5$ and (a) $b=\lambda$, (b) $b=\lambda/2$, (c) $b=0$. The other parameters are the same as in Fig.\ref{colourmap}. The swimmer trajectories are straight lines of lenght $\lambda$; only $1/10$ of the actual swimmer trajectory is shown. The beginning and the end of the swimmer and tracer trajectories are marked with empty circles. Arrows indicate the direction of motion. }
\label{loops}
\end{figure}

To better understand the nature of the integral in Eq.(\ref{A}), in Fig.\ref{colourmap} we plot its integrand $e^{2\xi}\,\tilde\Delta^2(\xi,\chi)$. Its general structure is similar to that of the "squirmer" model as calculated by Lin \emph{et al.} \cite{childress2011}. First we note that the integrand is significantly different from zero in a localised part of the domain and we expect the integral in Eq.(\ref{A}) to converge. Second, the asymmetric low-value part of the domain around $(0,0)$ is the consequence of the cut-off $c$. Finally, there are two distinct parts of the domain where the integrand is large: around $b/\lambda=0$ and $b/\lambda=1$. The origin of these peaks was explained by Lin \emph{et al.} \cite{childress2011} and is related to a finite value of $\lambda$. In Figs.\ref{loops}(a), (b) and (c) we plot individual scattering events for $a=1.5$ and $b=\lambda$, $b=\lambda/2$ and $b=0$, respectively. The scattering event in Fig.\ref{loops}(b) corresponds to a situation where the swimmers trajectory is symmetric with respect to the original position of the tracer. Such configuration is similar to scattering of a passive tracer by a swimmer with $\lambda=\infty$ which has been extensively studied\cite{Dunkel2010,childress2011,Childress2010,mino2013,pushkin2013jfm,pushkin2013prl} and it is well-understood that $\lambda=\infty$ scattering events result in closed or almost closed loop-like trajectories of the tracer particle. Indeed, as can be seen from Fig.\ref{loops}(b), for finite $\lambda$ the net displacement in a symmetric scattering is rather small. However, when the symmetry is broken, as in the event in Figs.\ref{loops}(a) and (c), the tracer particle follows only a part of an almost-closed loop similar to Fig.\ref{loops}(b), and the net displacement is significantly larger. Therefore, the effective diffusion coefficient of the tracer particle should strongly depend on the value of $\lambda$, as we will show in the next Section. 

To evaluate the integral in Eq.(\ref{A}), we sum the calculated values of the integrand in Fig.\ref{colourmap}, multiply it by the total area of the domain and divide by the number of the grid points used. The errors at the boundaries of the domain introduced by this method are of the order of the grid-spacing and are small since there the integrand is very small. Although Fig.\ref{colourmap} suggests that only a small portion of the $(\chi,\xi)$-domain contributes to the integral, we found that in order to obtain a value converged to the third significant figure, it was important to extend the integration domain to $[-30,30]\times[-30,30]$ with the grid-spacing $0.05$. Finally, we obtain
\begin{equation}
A = 3.75.
\label{Adrescher}
\end{equation}
This value is specific for the dipolar velocity field, Eq.(\ref{dipolar}), and our choice of parameters: $p=32$, $U=22$, $\lambda=10$ and $c=1$. In the following Section we show that while $A$ is a weak function of $p$, $U$ and $c$, it strongly depends on $\lambda$.

\section{Many-Particle Simulations}
\label{simulations}

The theoretical arguments presented in Section \ref{theory}, rely, intrinsically, on the assumption of sequential dynamics: each scattering event was assumed to take place independently from other scattering events. In reality, however, the tracer particle would never be able to complete a trajectory of the type shown in Fig.\ref{loops} since it is constantly being displaced in various directions by other swimmers. Intuitively, it follows that the resulting displacements are smaller than the full scattering trajectories as in Fig.\ref{loops}, and the effective diffusion coefficient is reduced from the estimate in Eqs.(\ref{main}) and (\ref{Adrescher}).  

To quantify the effect of simultaneous scattering by many swimmers, we perform simulations of self-propelled point-like dipoles and point-like tracers in a 3D box of size $L$. We keep exactly the same assumptions and the dynamics as in Section \ref{theory} with only one difference: a tracer particle is moved according to the velocity field created by all the swimmers at its position. We apply periodic boundary conditions for the swimmers and the tracer particles. We set $p=32$, $U=22$, $c=1$, $\lambda=10$ and $n=10^{-3}$ unless specified otherwise. The numerical protocol consists of running $40$ independent simulations for $100-150$ time units with $500$ tracers in each simulation. The resulting mean-squared displacements are averaged over all tracers and all realisations.

\begin{figure}[t]
\centering
\includegraphics[height=4.5cm]{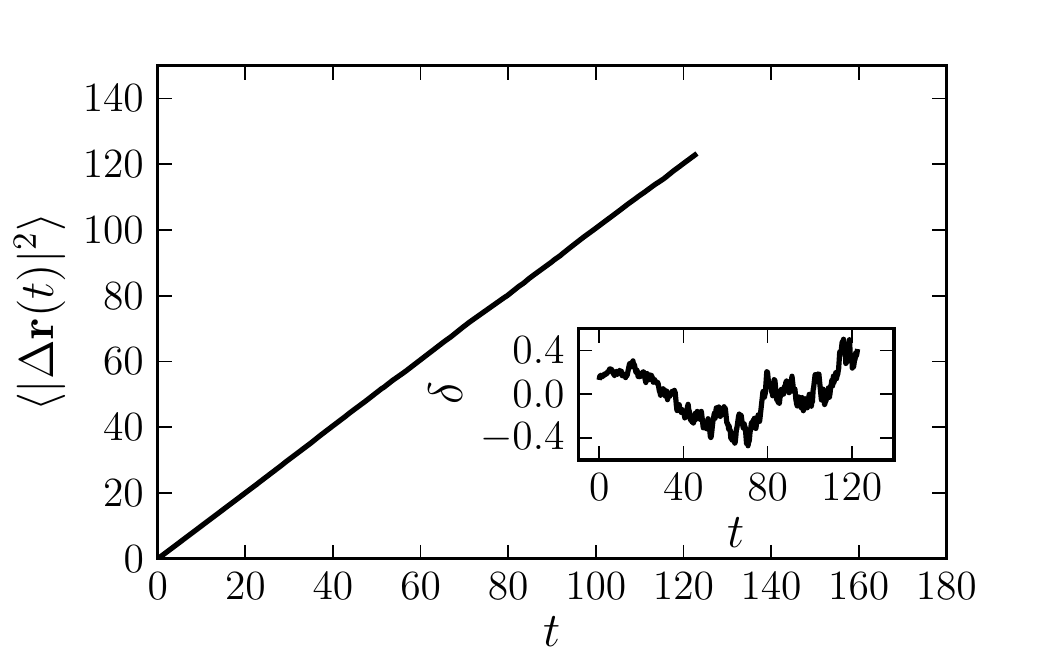}
\caption{The mean-squared displacement of a tracer particle as a function of time for the same parameters as in Fig.\ref{colourmap} and the number density of swimmers $n=10^{-3}$. The results are averaged over $500$ tracers. Inset: the difference $\delta=\langle |\Delta{\bf r}(t)|^2\rangle-6\,D\,t$ with $D=0.167$ obtained from the linear fit.}
\label{diffusive}
\end{figure}

In Fig.\ref{diffusive}, we show the averaged mean-squared displacement as a function of time for $L=200$ and $dt=10^{-3}$, and in the inset we plot the difference $\delta$ between the mean-squared displacement and $6\,D\,t$, where $D=0.167$ is determined from a linear fit. The difference $\delta$ is essentially a small-magnitude noise demonstrating a purely diffusive behaviour of the tracers. Previous experiments\cite{wu2000,Chantal2011} and simulations\cite{Underhill2008,Chantal2011} have also reported normal diffusion at long times in suspensions of self-propelled particles. Similar to these studies, we observe superdiffusive behaviour at short times, $t\le1$, which we did not study here. Using the measured value of $D$ and assuming the scaling from Eq.(\ref{main}), we obtain that $A=3.60$ for simultaneous scattering by many swimmers. As expected, this value is lower than the theoretical estimate Eq.(\ref{Adrescher}), although the difference might be smaller than a typical experimental error in determining $D$.

Our next goal is to assess the validity of the scaling in Eq.(\ref{main}). As we have mentioned above, the structure of Eq.(\ref{main}) implies that it captures the main dependence of the effective diffusion coefficient on $n$, $U$ and $p$, while $A$ is a numerical prefactor. Here we show that $A$ is in fact a function of the system parameters, and while some dependencies are residual, others are crucial. We are going to systematically change various parameters in our simulations and study their effect on $A$, which we define as the measured value of the diffusion coefficient $D$ divided by the main scaling term $n\,U\left(p/U\right)^2$ from Eq.(\ref{main}). We will call our main set of parameters ($p=32$, $U=22$, $c=1$, $\lambda=10$ and $n=10^{-3}$) the \emph{E. coli} parameters.

\begin{figure}[t]
\centering
\includegraphics[height=4.5cm]{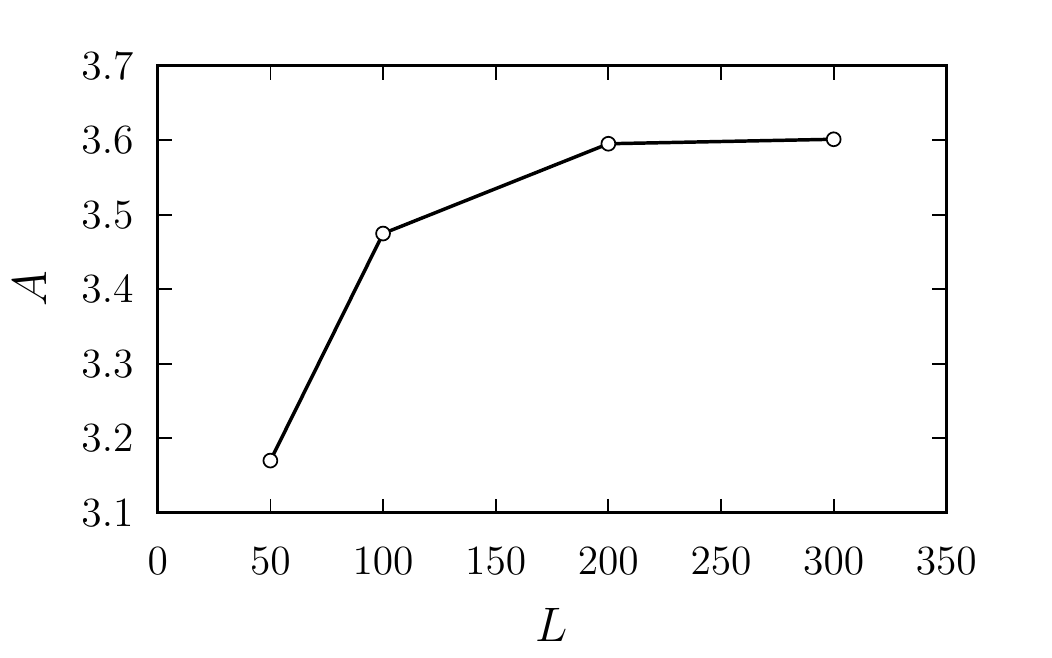}
\caption{Dependence of the prefactor $A$ on the system size for the \emph{E. coli} parameters. The number of the swimmers is adjusted to keep $n=10^{-3}$ for each $L$.}
\label{Lscaling}
\end{figure}

\begin{figure}[t]
\centering
\includegraphics[height=4.5cm]{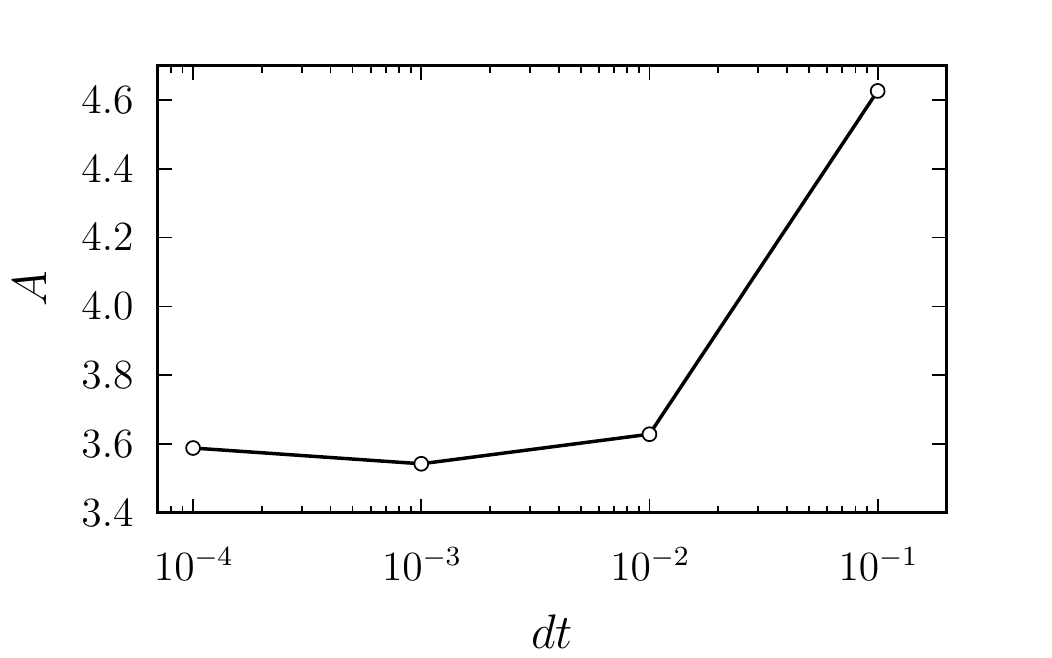}
\caption{Numerical convergence of the prefactor $A$ for various time-steps $dt$.}
\label{dtscaling}
\end{figure}

First, we determine the spatial and temporal accuracy needed to produce converged results. In Fig.\ref{Lscaling}, we plot $A$ for the \emph{E. coli} parameters and various system sizes $L$ where we have changed the number of swimmers accordingly to keep $n=10^{-3}$. We observe that the result has converged for $L=200$. Next, we study how $A$ changes with the timestep $dt$ for $L=200$. Fig.\ref{dtscaling} shows that a timestep in the range $10^{-4}-10^{-2}$ produces sufficiently accurately results. Therefore, in the following we set $L=200$ and $dt=10^{-3}$. 

\begin{figure*}
\centering
\includegraphics[height=3.7cm]{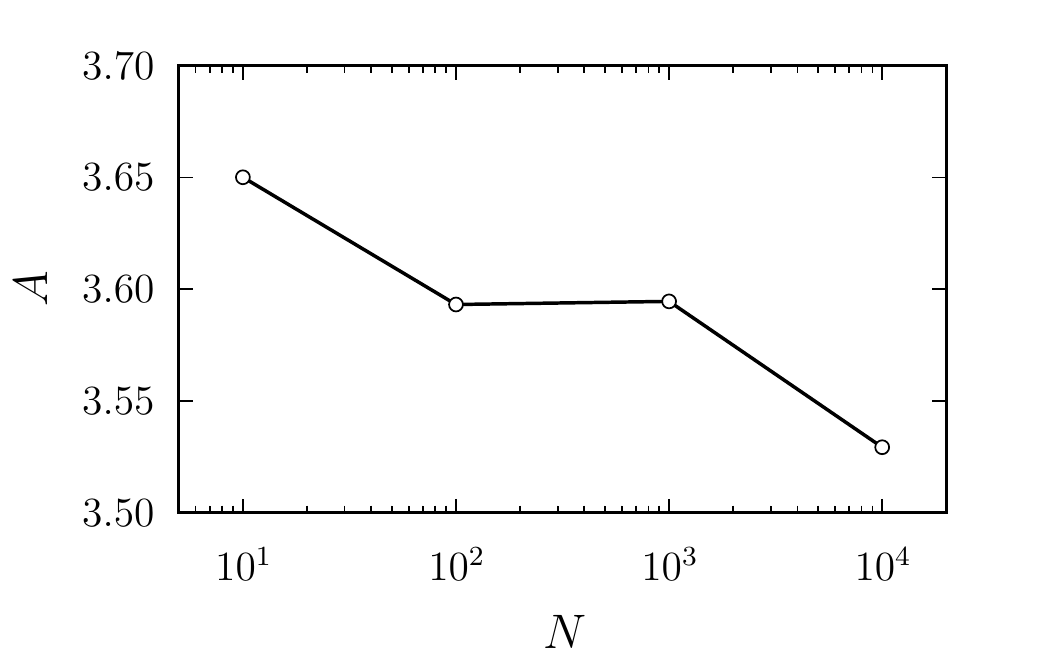}
\includegraphics[height=3.7cm]{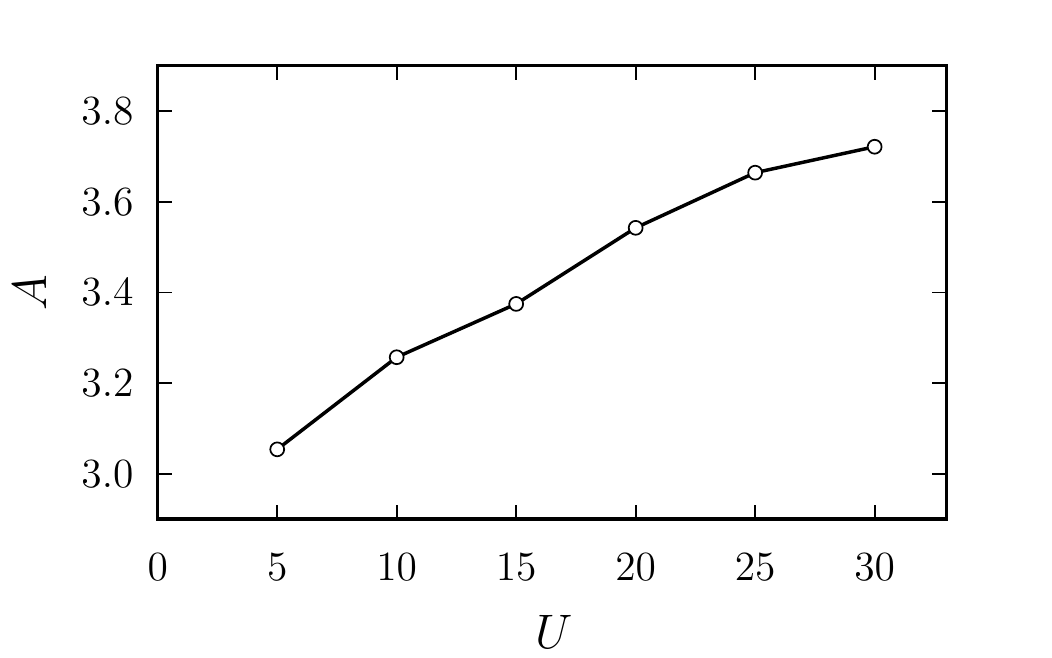}
\includegraphics[height=3.7cm]{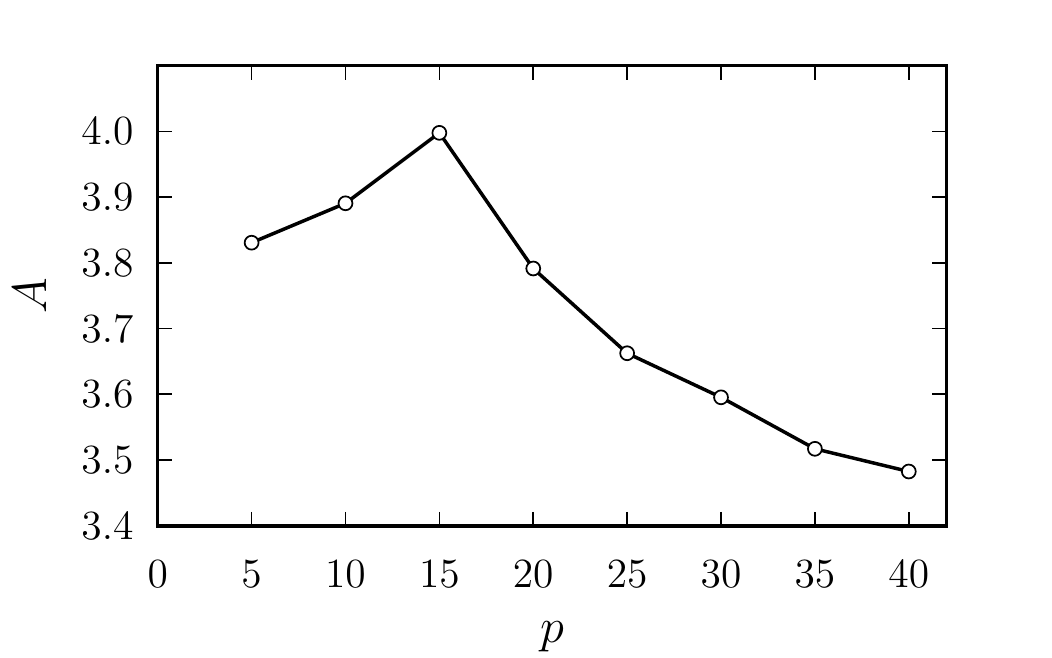}
\caption{Dependence of the prefactor $A$ on (left) the number of the swimmers $N$, (middle) their swimming speed $U$, and (right) their dipolar strength $p$. The other parameters are kept to be the same as our main \emph{E. coli} values.}
\label{mainscaling}
\end{figure*}

Next, we check the residual dependence of $A$ on $p$, $U$ and the number of swimmers $N$ that determines the number density $n=N/L^3$. Fig.\ref{mainscaling} demonstrates that $A$ is rather insensitive to the number density and is a mild function of $U$ and $p$ in the range of values relevant for \emph{E. coli} bacteria. It implies that while Eqs.(\ref{main}) and (\ref{Adrescher}) are sufficient to give the correct order of magnitude, in order to quantitatively predict the effective diffusion coefficient, $A$ should be calculated using $U$ and $p$ for the particular bacterial suspension in question. 
To illustrate this point, we consider the experiments by Jepson \emph{et al.}\cite{alys2013} who measured the enhanced diffusivity of tracers in 3D \emph{E. coli} suspensions to be $\Delta D/\left(n\,U\right) = 7\pm0.4\mu m^4$ for bacteria swimming with the average speed $U=15\mu m/s$. Using Eqs.(\ref{main}) and (\ref{Adrescher}), we obtain $\Delta D/\left(n\,U\right)= 7.93\mu m^4$ which is slightly different from the measured value. However, the value of $A=3.75$ in Eq.(\ref{Adrescher}) was calculated for $U=22$ as in Drescher \emph{et al.} \cite{Drescher2011}. If we use $U=15$, which is the value relevant for the particular strain of Jepson \emph{et al.}\cite{alys2013}, and perform integration in Eq.(\ref{A}) on the domain $[-10,4]\times[-10,10]$ with the grid-spacing $0.02$ corresponding to their experimental conditions, we obtain $\Delta D/\left(n\,U\right)= 7.24\mu m^4$ which is within the error bars of the experimental values.

\begin{figure}[t]
\centering
\includegraphics[height=4.5cm]{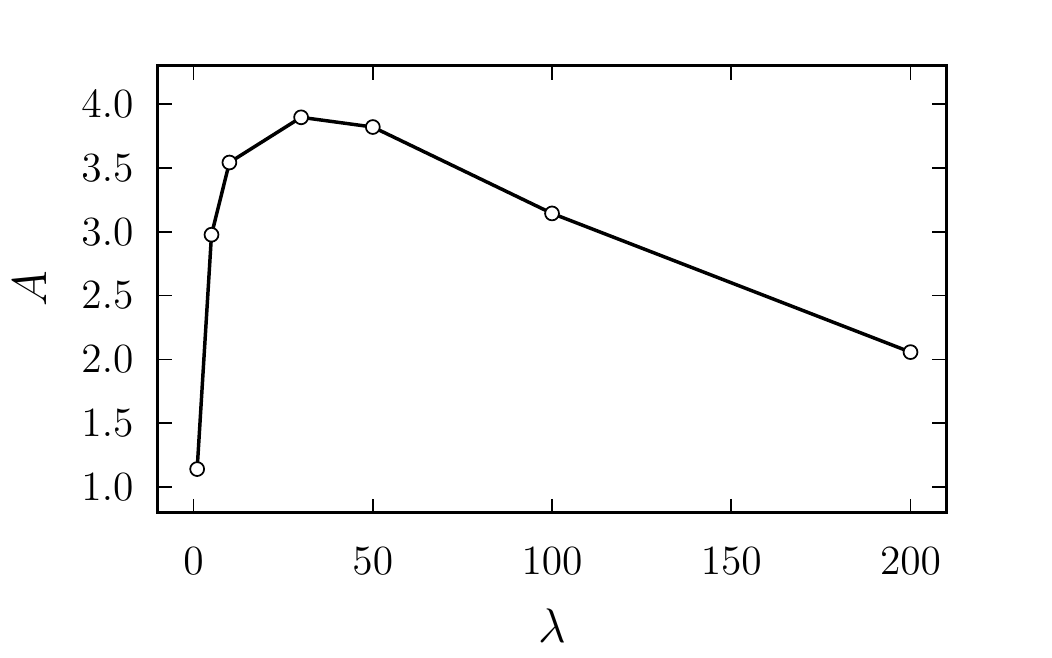}
\caption{Prefactor $A$ as a function of the run length $\lambda$.}
\label{RLscaling}
\end{figure}

\begin{figure}[t]
\centering
\includegraphics[height=4.5cm]{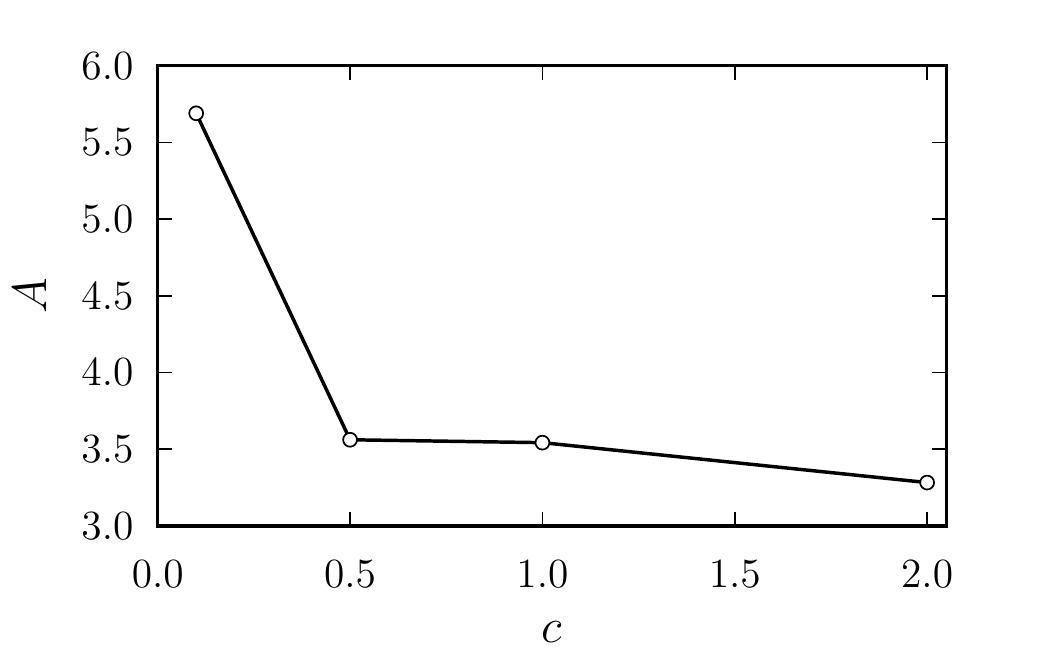}
\caption{Prefactor $A$ as a function of the short-range cut-off $c$.}
\label{Cscaling}
\end{figure}

Now, we turn our attention to the dependence of $A$ on the run length $\lambda$. As has been argued above, at infinite $\lambda$ the net displacement of the tracer particle during each scattering event is minute, and the effective diffusion coefficient is small. Indeed, as Fig.\ref{RLscaling} shows, $A$ depends strongly on $\lambda$, decreasing rapidly for large $\lambda$. Interestingly, the maximum values of $A$ are observed for the biologically relevant range of the run lengths. Finally, in Fig.\ref{Cscaling} we demonstrate how $A$ depends on the cut-off distance $c$ that we have introduced in order to mimic the finite size of the bacteria and the tracers. As expected, the exact value of the cut-off distance has only a small effect on the diffusion coefficient unless $c$ is rather small so that the tracer can get quite close to a swimmer's core. In that case, the main contribution to the effective diffusion is not advection by the swimmers far away, as we have assumed until now, but instead the transport in the wake of a single swimmer\cite{childress2011,pushkin2013jfm,pushkin2013prl}. As we will show below, this mechanism depends strongly on the details of the near-field of the swimmer that we have been neglecting throughout this paper. However, for large enough values of $c$, typically $c>0.5$ which corresponds to the scattering of micron-size particles by swimming bacteria, the tracer particle cannot approach the core of the swimmer close enough, and the details of the near-field are largely irrelevant.
  
\begin{figure}[t]
\centering
\includegraphics[height=4.5cm]{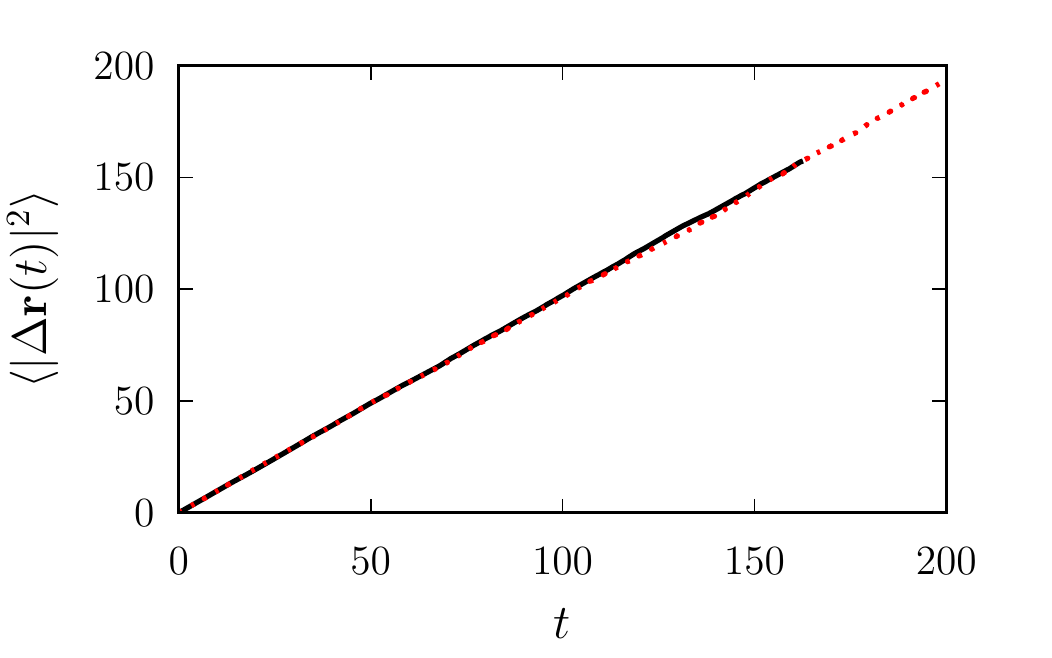}
\caption{Mean-squared displacement of a tracer particle subject to random noise and hydrodynamic interactions with the swimmers for the \emph{E. coli} parameters and $L=100$. The strength of the noise is chosen to give the tracer's diffusion coefficient $D_{th}=0.2$ in the absence of the swimmers. Black solid line: no thermal diffusion. Red dotted line: the difference between the observed mean-squared displacement with the noise and the mean-squared displacement due to thermal diffusion $\langle|\Delta{\bf r}(t)|^2\rangle-6\,D_{th}\,t$.}
\label{brownian}
\end{figure}  
 
Finally we comment on the effect of Brownian motion on the diffusivity of the tracer particles. It is \emph{a priori} unclear whether the effects of thermal diffusion and the effective diffusion due to advection by the swimmers simply add up or whether there is an interaction between these two effects. It is expected that the presence of thermal diffusion significantly alters the outcome of a single scattering event since the tracer particle would be constantly moved away from the scattering loop that it would trace otherwise\cite{Dunkel2010}. However, the accumulated effect of the thermal noise on the long-time displacement of the tracer is unclear. Therefore, we perform simulations where tracers are also subject to thermal diffusion with the diffusion coefficient chosen to be $D_{th}=0.2$, which, in our units, corresponds to the thermal diffusion coefficient of a micrometer-size particle in water at room temperature. We use $L=100$ to speed up simulations. In Fig.\ref{brownian} we plot the mean-squared displacement of the tracer particle $\langle|\Delta{\bf r}(t)|^2\rangle$ in the case of no thermal diffusion (black solid line) and $\langle|\Delta{\bf r}(t)|^2\rangle-6\,D_{th}\,t$ in the case of the thermal diffusion (red dotted line). It is clear that the two effects are additive and, therefore, the enhanced diffusion by swimming agents can be studied in isolation from the thermal effects.

\section{Enhanced diffusion in 2D}
\label{2Dtheory}

While the problem of enhanced diffusivity in 3D bacterial suspensions is relevant for large volumes like bacteria in oceans\cite{Katija2012}, typical laboratory experiments deal with rather small samples where bacteria can swim from one confining surface to another in a relatively short time. In small volumes, bacteria tend to accumulate next to confining surfaces \cite{berke2008,tang2009} making it difficult to study enhanced diffusivity in the bulk. In fact, most of the experimental studies on enhanced diffusion in bacterial suspensions were performed in 2D or quasi-2D systems\cite{wu2000,mino2011,mino2013,Chantal2011,kurtuldu2011}. Therefore, we repeat the calculation of Section \ref{theory} to estimate the effective diffusion coefficient of a tracer in a 2D bath of swimming bacteria. In this Section we assume that the dynamics of both the swimmers and the tracer are confined to a plane parallel to a flat solid wall. We refer to this plane as the \emph{swimming plane}. The distance between the swimming plane and the wall is $h$ and the velocity of the fluid is assumed to satisfy the no-slip boundary condition at the wall. We choose a Cartesian coordinate system with the $z$-direction perpendicular to both the wall and the swimming plane and set the wall at $z=0$. 

Similar to free swimming, the velocity field produced by a swimming bacterium next to a solid wall is well-described by a force-dipole \cite{Drescher2011}. Its analytical form is, essentially, given by Eq.(\ref{dipolar}) modified to satisfy the no-slip boundary condition at the wall and can be readily obtained from a well-known expression for the velocity field ${\bf u}_s$ at the position ${\bf r}$ produced by a point-force $\bf f$ applied to the fluid at a distance $h$ from the wall\cite{kimkarrila,Spagnolie2012}
\begin{eqnarray}
&&{\bf u}_s\left( {\bf r}\right) = \frac{f}{8\pi\eta} \left[ {\bf e}\left( \frac{1}{r} - \frac{1}{R} - \frac{2h^2}{R^3}\right) + \right. \nonumber \\
&&\qquad\qquad\qquad \left. + {\bf r} \left( {\bf r}\cdot{\bf e}\right) \left( \frac{1}{r^3} - \frac{1}{R^3} + \frac{6h^2}{R^5}\right) \right].
\end{eqnarray}
Here, $f$ is the magnitude of the force, $\bf e$ is the unit vector in the direction of the force, and $\eta$ is the viscosity of the fluid; $\bf r$ and $\bf e$ are assumed to lie in the swimming plane, $r=|{\bf r}|$ and $R=\sqrt{r^2+4h^2}$. The dipolar velocity field produced by two point-like forces of the same magnitude and opposite directions at a distance $h$ from the wall is then given by
\begin{eqnarray}
{\bf u}\left( {\bf r}\right) = p\,{\bf r}\left[ \frac{1}{r^3}\left( 3\cos^2{\theta} - 1\right) - \frac{1}{R^3}\left( 3\frac{r^2}{R^2}\cos^2{\theta} - 1\right) \right. \nonumber \\
\left. + \frac{6h^2}{R^5}\left( 5\frac{r^2}{R^2}\cos^2{\theta} - 1\right) \right] - 12\frac{h^2}{R^5}\left( {\bf r}\cdot{\bf e}\right){\bf e},
\label{2Dvelocity}
\end{eqnarray}
where $\bf e$ coincides with the swimming direction of the resulting force dipole, $\theta$ is the angle between $\bf e$ and $\bf r$, $p=F \delta/8\pi\eta$, and $\delta$ is the length of the dipole. In the limit $h\rightarrow\infty$, Eq.(\ref{2Dvelocity}) reduces to the 3D dipolar field given by Eq.(\ref{dipolar}). 

\begin{figure}[t]
\begin{center}
\includegraphics[width=9.cm]{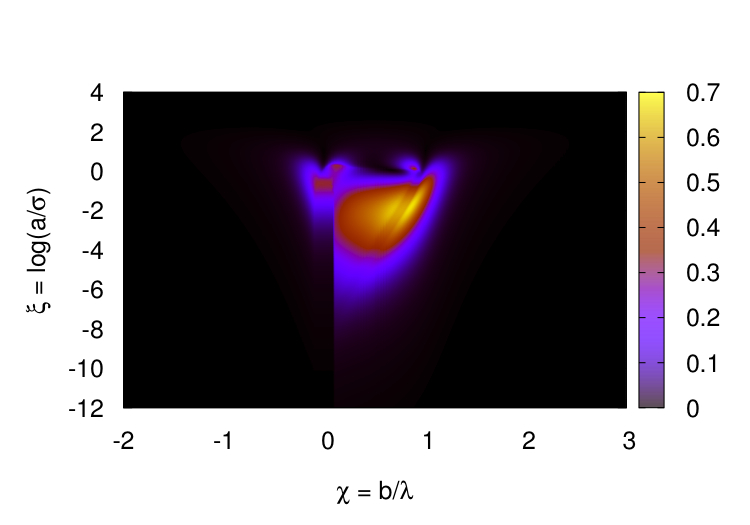}
\end{center}
\caption{The value of the integrand $e^{\xi}\,\tilde\Delta^2(\xi,\chi,h)$ from Eq.(\ref{A2D}) for $h=1$. The other parameters are the same as in Fig.\ref{colourmap}.}
\label{colourmap2D}
\end{figure}

The 2D analogue of Eq.(\ref{int_original}) was obtained by Lin \emph{et al.} \cite{childress2011} and reads
\begin{equation}
\langle |\Delta{\bf r}(t)|^2\rangle =2\,n_{2D} \frac{U t}{\lambda}  \int_{0}^{\infty}da \int_{-\infty}^{\infty} db\,\Delta^2(a,b) \equiv 4\,D_{2D}\,t,
\label{int2D}
\end{equation}
where $n_{2D}$ is the surface number density and $\Delta$ is the displacement of the tracer particle in the swimming plane parametrised as before by $a$ and $b$ from Fig.\ref{event}. Using the same normalisation as in Section \ref{theory}, we obtain
\begin{equation}
D_{2D}(h) = A_{2D}(h)\,n_{2D}\,U \left(\frac{p}{U}\right)^{3/2},
\label{main2D}
\end{equation}
where
\begin{equation}
A_{2D}(h) = \frac{1}{2}\int_{-\infty}^{\infty}d\xi \int_{-\infty}^{\infty}d\chi\,e^{\xi}\,\tilde\Delta^2(\xi,\chi,h).
\label{A2D}
\end{equation}
In Fig.\ref{colourmap2D} we plot the value of the integrand $e^{\xi}\,\tilde\Delta^2(\xi,\chi,h)$ for $p=32$, $U=22$, $c=1$, $dt=0.001$, $\lambda=10$ and $h=1$. It is clear that the 2D integrand is more localised in space and is weaker than its 3D counterpart, Fig.\ref{colourmap}, which can be attributed to the weakening of the velocity field by the presence of a solid boundary. To obtain $A_{2D}(h)$, we perform numerical integration in Eq.(\ref{A2D}) on the domain $(\chi,\xi) = [-5,5]\times[-12,5]$ with the grid-spacing $0.05$. We have checked that this resolution ensures numerical convergence of the integral to the third significant figure. 

\begin{figure}[t]
\centering
\includegraphics[height=4.5cm]{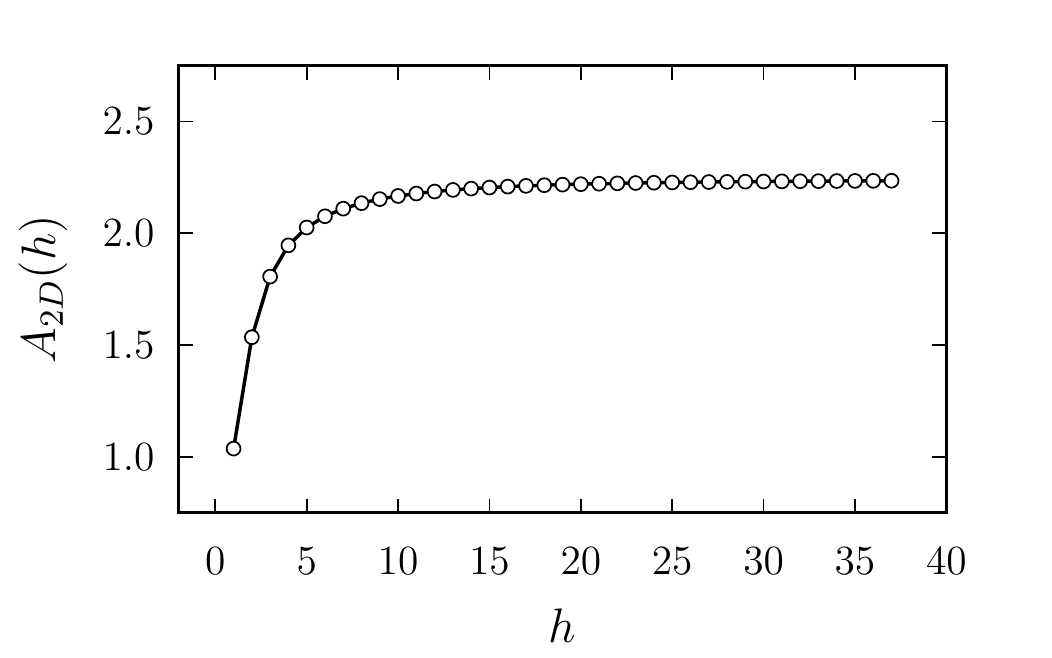}
\caption{Prefactor $A_{2D}$ in two dimensions as a function of the distance to the wall $h$ for the \emph{E. coli} parameters.}
\label{A2Dfig}
\end{figure}

Our main result for the effective diffusion coefficient in a 2D plane adjacent to a solid wall are given by Eq.(\ref{main2D}) and Fig.\ref{A2Dfig}. We observe that the diffusion is significantly suppressed closed to the wall and then saturates to a constant value far away from the wall. Since in our simulations the length is measured in micrometers, Fig.\ref{A2Dfig} predicts that the effect of the wall vanishes at distances around $10\mu m$ for the \emph{E. coli} parameters as measured by Drescher \emph{et al.} \cite{Drescher2011}. We also note that the saturated value of the diffusion coefficient is different from the 3D case:  if we "cut" from a 3D suspension with number density $n$ a slice of thickness $\sigma=\sqrt{p/U}$ and introduce an effective 2D number density in the slice, $n\sqrt{p/U}$, Eq.\ref{main} predicts a diffusion coefficient which is almost twice larger than the saturated value in Fig.\ref{A2Dfig}. This clearly demonstrates the effect of the dimensionality on $D$.

We compare our prediction, Eqs.(\ref{main2D}) and (\ref{A2D}), with the experimental data on quasi-2D bacterial suspensions by Mi\~no \emph{et al.}\cite{mino2013}. In their experiments, Mi\~no \emph{et al.}\cite{mino2013} followed buoyant tracers and bacteria in a $5\mu m$-deep layer next to a solid wall. For the value of the active flux $n\,U = 0.01 bact/(\mu m^2 s)$ they have measured the enhanced diffusion coefficient $D \approx 0.14 \mu m^2/s$ (see Fig.8 in Mi\~no \emph{et al.}\cite{mino2013}). There, the 3D number density $n$ was determined by counting all the bacteria in the field of view of the microscope and assuming that they are uniformly distributed across the $5\mu m$-thick layer. First we note that the 3D prediction, Eqs.(\ref{main}) and (\ref{A}), yield $D\approx 0.08 \mu m^2/s$, which is significantly lower than the measured value. Next, we use Eq.(\ref{main2D}) and Fig.\ref{A2Dfig} to produce a quasi-2D estimate of the enhanced diffusion coefficient $\bar{D}_{2D}$. We average $A_{2D}(h)$ from $c/2$ to $H=5$ in our units to produce
\begin{eqnarray}
&&\bar{D}_{2D} = n_{2D}\,U \left(\frac{p}{U}\right)^{3/2} \frac{1}{H} \int_{c/2}^{H}A_{2D}(h)dh \nonumber\\
&&\qquad\qquad = 7.26\,n\,U \left(\frac{p}{U}\right)^{3/2},
\label{Daverage2D}
\end{eqnarray}
where we have introduced the 3D number density $n=n_{2D}/H$. For $n\,U = 0.01 bact/(\mu m^2\cdot s)$, as in Mi\~no \emph{et al.}\cite{mino2013}, and using $p/U = 32/22 \mu m^2$, we obtain $\bar{D}_{2D}=0.13\mu m^2/s$, which is significantly closer to the measured value than the 3D prediction mentioned above. There are several sources of potential discrepancies here. First, the geometric factor $\sigma=\sqrt{p/U}$ might be different for the strain of \emph{E. coli} used in the experiments by Mi\~no \emph{et al.}\cite{mino2013} than the value measured by Drescher \emph{et al.} \cite{Drescher2011} used here. Second, the distribution of the bacteria next to the wall is most likely very inhomogeneous with significantly higher concentrations in the immediate vicinity of the wall\cite{berke2008,tang2009}. Finally, in our estimate we have averaged over swimming planes at various distances from the wall neglecting bacterial motion in the direction perpendicular to the wall that is limited but non-zero. Nevertheless, our estimate of $\bar{D}_{2D}$ is remarkably close to the experimental value questioning the recent statement by Pushkin and Yeomans\cite{pushkin2013prl} that in 2D enhanced diffusion is dominated by the entrainment mechanism. Here we have demonstrated that a simple dipolar field produced by bacteria is sufficient to produce a good estimate of the enhanced diffusivity in both 2D and 3D suspensions.

\section{Discussion}
\label{discussion}

The main result of our paper is the prediction for the effective diffusion coefficient, Eq.(\ref{main}) in 3D and Eq.(\ref{main2D}) in 2D. First of all, we have confirmed the previously reported scaling of the effective diffusivity with the active flux. The rest of the scaling factor, $\left(p/U\right)^2$ in 3D and $\left(p/U\right)^{3/2}$ in 2D, can be understood by observing that the dipolar field, Eq.(\ref{dipolar}), can be generated by two non-interacting spheres moving with the speed $U$ along the line of their separation\cite{baskaran2009}. In this case, $p\sim Ul_1 l_2$, where $l_1$ is the typical size of the spheres, and $l_2$ is their separation. For \emph{E. coli}, a natural scale for $l_1$ is the radius of the bacterial body's cross-section. The separation $l_2$ can be thought of as the distance between two point forces applied to the fluid, one - at the centre of the bacterial body, the other - at the point where the force generated by the flagellar bundle is maximum. The propulsion force builds up from the free end of the flagellar bundle towards the cell body and reaches its maximum just outside of the hydrodynamic wake created by the moving body. Estimating the size of the wake to be of the same size as the bacterial body, we obtain $l_2\sim 3 l$, where $l$ is the half-length of the body. Using the typical \emph{E. coli} values,  $l_1=0.5\mu m$ and $l=1\mu m$, we obtain $p/U\sim 1.5\mu m^2$, which is close to the value $p/U=1.45\mu m^2$ measured by Drescher \emph{et al.} \cite{Drescher2011}.
This shows that $\sqrt{p/U}$ should not be thought of as a single lengthscale but is instead a measure of the geometrical asymmetry of the swimmers.

Our second observation is based on the results of many-particle simulations in Section \ref{simulations} where we have shown that the integral $A$ is a function of the properties of the swimmers. While Eqs.(\ref{main}) and (\ref{A}) are sufficient to predict the order of magnitude of the effective diffusion coefficient, our results suggest that in order to make a quantitative prediction for a particular suspension, one should use the value of $A$ which is specific to that suspension. Thus Fig.\ref{Lscaling} demonstrates that our estimate of $A$, Eq.(\ref{A}), is only valid for samples with the smallest dimension larger than $200\mu m$, and for smaller samples one should use a smaller value of $A$. This is related to the long-range nature of the dipolar field, Eq.(\ref{dipolar}): in small samples a significant contribution to the effective diffusion coefficient from bacteria far away would be absent compared to very large samples. 

The results of Section \ref{simulations} suggest that to predict quantitatively the value of the diffusion coefficient, it is important to use the values of the parameters specific to the particular bacterium in question. Fig.\ref{mainscaling} demonstrates that the coefficient $A$ is sensitive to the values of $U$ and $p$. In Section \ref{simulations} we illustrated this point by showing that the experimental results by Jepson \emph{et al.}\cite{alys2013} can only be accurately reproduced when using $A$ for the swimming speed observed in their experiments. In the similar fashion, the estimate of the enhanced diffusivity of "squirmers" by Lin \emph{et al.} \cite{childress2011} is likely to be somewhat inaccurate since they have assumed that the precise value of $U$ was not important and it was set to unity while calculating $A$. Lin \emph{et al.} \cite{childress2011} then attempted to reproduce the value of the enhanced diffusivity of \emph{Chlamydomonas reinhardtii} measured by Leptos \emph{et al.}\cite{Leptos2009} by using this value of $A$ at $U=1$ and a different value of $U$ in the scaling prefactor. Fig.\ref{mainscaling} implies that the same value of $U$ should have been used in both the scaling expression and the prefactor $A$ to achieve an accurate result. While  it might seem that here we are addressing rather small differences, the potential discrepancies in using a result for one type of swimmers for another organism can be large, as suggested by Fig.\ref{mainscaling}.

The most essential parameter that influences the effective diffusion coefficient is the run length of the bacteria $\lambda$. In Fig.\ref{RLscaling} we show that although Eqs.(\ref{main}) and (\ref{A}) do not seem to contain $\lambda$, in fact $A$ is a very strong function of the bacterial run length. The mechanism of this dependence was outlined by Lin \emph{et al.} \cite{childress2011} and explained in detail in Section \ref{theory}. Essentially, for very long run lengths tracers perform almost closed loops with small net displacements, while for short run lengths both the total and net displacements are small. For intermediate values of $\lambda$, the net displacements are significant resulting in a large effective diffusion coefficient, as can be seen in Fig.\ref{RLscaling}. One implication of this result is that bacterial mutants that tumble a lot and smooth swimmers that only deviate from a straight path due to rotational diffusion should result in significantly smaller diffusivity of tracer particles than the wild-type \emph{E. coli}. The first part of this prediction is supported qualitatively by the data from Kim and Breuer\cite{Kim2004} who observed that the diffusion coefficient of \emph{Dextran} molecules in  solutions of tumbly \emph{E. coli} was about twice smaller than in solutions of wild-type bacteria.

Recently, Pushkin and Yeomans proposed\cite{pushkin2013prl} that in addition to the far-field advection mechanism discussed here, there is an additional contribution to the enhanced diffusivity due to the entrainment of tracers by the bacteria passing in their close vicinity. They argued that this contribution should be especially relevant for quasi-2D suspensions. The excellent numerical agreement between our estimates and the measured 3D enhanced diffusivity by Jepson \emph{et al.}\cite{alys2013} suggests that the entrainment mechanism is irrelevant in 3D for large tracers (see below). Similarly, a naive generalisation of our purely 2D prediction, Eq.(\ref{main2D}), to the quasi-2D case, Eq.(\ref{Daverage2D}), is in a good agreement with the values measured by Mi\~no \emph{et al.}\cite{mino2013} accounting for about $90\%$ of the measured value. While the other $10\%$ might be attributed to either the entrainment and/or experimental error, we note here that our simple theory based on the far-field hydrodynamics is quite sufficient to predict the enhanced diffusion coefficient also in 2D. 

\begin{figure}[t]
\centering
\includegraphics[height=4.5cm]{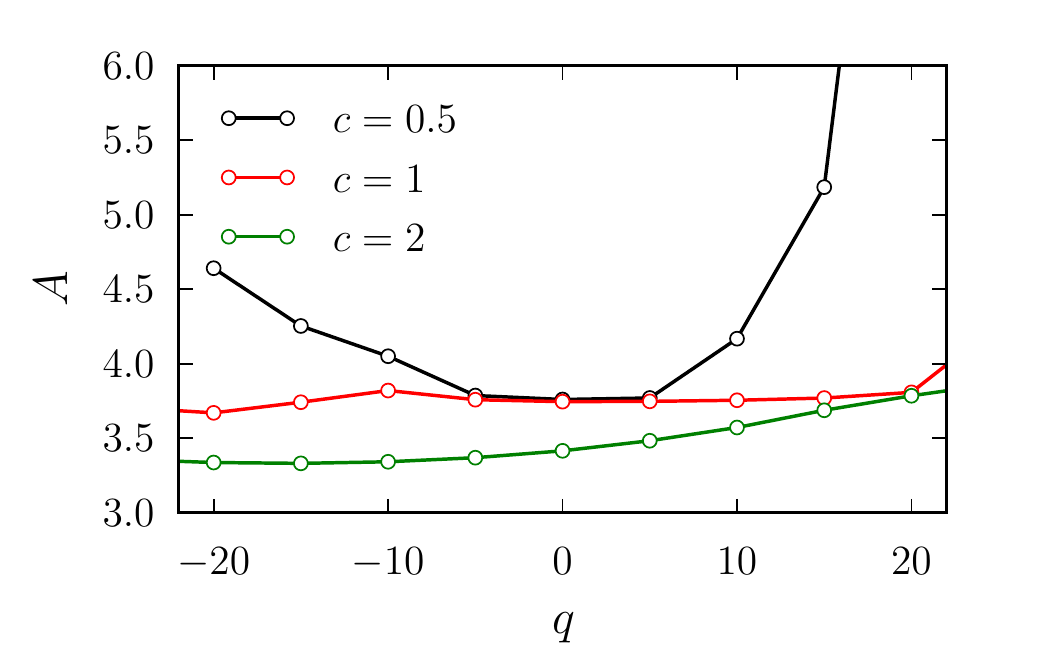}
\caption{Prefactor $A$ for swimmers with the dipolar and quadrupolar velocity fields, Eqs.(\ref{dipolar}) and (\ref{quadrupolar}) as a function of the quadrupolar strength for the \emph{E.coli} parameters.}
\label{Aquad}
\end{figure}

Calculations presented in this paper rely on the assumption that the velocity field produced by the bacteria is purely dipolar. In reality, however, bacteria create additional short-ranged fields in the vicinity of their bodies as demonstrated by Drescher \emph{et al.} \cite{Drescher2011}. In order to estimate the importance of the bacterial near-field, we repeat here the calculation from Section \ref{theory} where in addition to the velocity field Eq.(\ref{dipolar}), we also consider a quadrupolar contribution\cite{Spagnolie2012}
\begin{equation}
{\bf u}_Q\left( {\bf r}\right) = q \left[  \frac{3\,{\bf r}\cos{\theta}}{r^4}  \left( 5\cos^2{\theta} - 3\right) - \frac{{\bf e}}{r^3}\left(3\cos^2{\theta} - 1\right)\right],
\label{quadrupolar}
\end{equation}
where $q$ sets the strength of the field. We calculate the coefficient $A$ as the function of the quadrupolar strength $q$ for three values of the short-range cut-off $c$. The rest of the parameters are kept the same as in Section \ref{theory}. We did not find any measurements of the quadrupolar strength $q$ for \emph{E. coli} bacteria, but Drescher \emph{et al.} \cite{Drescher2011} have plotted the difference between the bacterial velocity field and the leading dipolar contribution, allowing us to estimate $|q|\sim 20 \mu m^4/s$. The sign of $q$ could not be determined since Drescher \emph{et al.} \cite{Drescher2011} have only plotted the absolute value of the velocity. As can be seen from Fig.\ref{Aquad}, the value of $A$ is largely insensitive to the quadrupolar strength $q$ as long as the cut-off value is large enough. Since we interpret the cut-off as the shortest separation between the centres of a bacterium and a tracer, Fig.\ref{Aquad} suggests that for tracers comparable in size to the bacteria, the main contribution to the diffusion coefficient comes from the dipolar far-field created by the bacteria. In this situation, neither the precise details nor the strength of the near-field really matters. On the contrary, for small tracers that can come much closer to the core of a swimmer, the diffusion coefficient is dominated by the near-field and is significantly larger than the diffusion coefficient for larger tracers. Kim and Breuer\cite{Kim2004} studied the enhanced diffusivity of small Dextran molecules in dilute solutions of wild-type \emph{E. coli}, and for $n=10^{-3}\mu m^{-3}$ reported $D\sim30\mu m^2/s$, while Jepson \emph{et al.}\cite{alys2013} reported $D\sim0.1\mu m^2/s$ at the same concentration when the role of the tracers was played by dead bacteria. Our results suggest an explanation of this discrepancy based on the difference in the tracer size, as shown in Fig.\ref{Aquad}. It is also possible that for small molecules in 3D the entrainment mechanism becomes important, as discussed by Pushkin \emph{et al.}\cite{pushkin2013jfm,pushkin2013prl}.

Several recent studies \cite{Short2006,Michelin2011} indicated that swimming of microorganisms is optimised with respect to swimming efficiency, chemotactic strategies \emph{etc.} This work suggests an intriguing possibility that, at least in \emph{E. coli}, the parameters of bacterial kinematics might have co-evolved to optimise enhanced diffusivity of passive particles or droplets that can either be a source of organic material or oxygen. Fig.\ref{RLscaling} shows that the typical run length of the wild-type \emph{E. coli} bacteria are close to the values that optimise the effective diffusion of tracer particles in dilute suspensions. While, most certainly, the main criterion of selecting particular values of bacterial swimming parameters is to optimise chemotaxis, we speculate here that simultaneous optimisation of the enhanced diffusivity of passive particles might provide additional evolutionary advantages for bacteria.

\section*{Acknowledgment}
AM was funded by the EPSRC (EP/I004262/1). We gratefully acknowledge discussions with Mike Cates, Alys Jepson, Vincent A. Martinez, Wilson Poon and Mitya Pushkin. This work has made use of the resources provided by the Edinburgh Compute and Data Facility (ECDF) (http://www.ecdf.ed.ac.uk/).

\footnotesize{
\bibliography{enhanced_diffusion}
\bibliographystyle{rsc}
}

\end{document}